
\hsize=6.5truein
\hoffset=.3truein
\vsize=8.9truein
\voffset=.1truein
\font\twelverm=cmr10 scaled 1200    \font\twelvei=cmmi10 scaled 1200
\font\twelvesy=cmsy10 scaled 1200   \font\twelveex=cmex10 scaled 1200
\font\twelvebf=cmbx10 scaled 1200   \font\twelvesl=cmsl10 scaled 1200
\font\twelvett=cmtt10 scaled 1200   \font\twelveit=cmti10 scaled 1200
\skewchar\twelvei='177   \skewchar\twelvesy='60
\def\twelvepoint{\normalbaselineskip=14pt
  \abovedisplayskip 12.4pt plus 3pt minus 9pt
  \belowdisplayskip 12.4pt plus 3pt minus 9pt
  \abovedisplayshortskip 0pt plus 3pt
  \belowdisplayshortskip 7.2pt plus 3pt minus 4pt
  \smallskipamount=3.6pt plus1.2pt minus1.2pt
  \medskipamount=7.2pt plus2.4pt minus2.4pt
  \bigskipamount=14.4pt plus4.8pt minus4.8pt
  \def\rm{\fam0\twelverm}          \def\it{\fam\itfam\twelveit}%
  \def\sl{\fam\slfam\twelvesl}     \def\bf{\fam\bffam\twelvebf}%
  \def\mit{\fam 1}                 \def\cal{\fam 2}%
  \def\tt{\twelvett}
  \textfont0=\twelverm   \scriptfont0=\tenrm   \scriptscriptfont0=\sevenrm
  \textfont1=\twelvei    \scriptfont1=\teni    \scriptscriptfont1=\seveni
  \textfont2=\twelvesy   \scriptfont2=\tensy   \scriptscriptfont2=\sevensy
  \textfont3=\twelveex   \scriptfont3=\twelveex  \scriptscriptfont3=\twelveex
  \textfont\itfam=\twelveit
  \textfont\slfam=\twelvesl
  \textfont\bffam=\twelvebf \scriptfont\bffam=\tenbf
  \scriptscriptfont\bffam=\sevenbf
  \normalbaselines\rm}

\def\beginlinemode{\endmode
  \begingroup\parskip=0pt \obeylines\def\\{\par}\def\endmode{\par\endgroup}}
\def\beginparmode{\endmode
  \begingroup \def\endmode{\par\endgroup}}
\let\endmode=\par
{\obeylines\gdef\
{}}
\def\singlespace{\baselineskip=\normalbaselineskip}
\def\oneandahalfspace{\baselineskip=\normalbaselineskip
  \multiply\baselineskip by 3 \divide\baselineskip by 2}
\def\doublespace{\baselineskip=\normalbaselineskip \multiply\baselineskip by 2}
\newcount\firstpageno
\firstpageno=2
\footline={\ifnum\pageno<\firstpageno{\hfil}\else{\hfil\twelverm\folio\hfil}\fi}
\let\rawfootnote=\footnote              
\def\footnote#1#2{{\rm\singlespace\parindent=0pt\rawfootnote{#1}{#2}}}
\def\raggedcenter{\leftskip=2em plus 12em \rightskip=\leftskip
  \parindent=0pt \parfillskip=0pt \spaceskip=.3333em \xspaceskip=.5em
  \pretolerance=9999 \tolerance=9999
  \hyphenpenalty=9999 \exhyphenpenalty=9999 }
\parskip=\medskipamount
\twelvepoint            
\overfullrule=0pt       
\def\author                     
  {\vskip 3pt plus 0.2fill \beginlinemode
   \singlespace \raggedcenter \twelvesc}
\def\affil                      
  {\vskip 3pt plus 0.1fill \beginlinemode
   \oneandahalfspace \raggedcenter \sl}
\def\abstract                   
  {\vskip 3pt plus 0.3fill \beginparmode
   \doublespace \narrower \noindent ABSTRACT: }
\def\endtitlepage               
  {\endpage                     
   \body}
\def\body                       
  {\beginparmode}               

\def\subhead#1{                 
  \vskip 0.1truein             
  {\raggedcenter #1 \par}
   \nobreak\vskip 0.1truein\nobreak}
\def\refto#1{$|{#1}$}           
\def\references                 
  {\subhead{References}         
   \beginparmode
   \frenchspacing \parindent=0pt \leftskip=1truecm
   \parskip=8pt plus 3pt \everypar{\hangindent=\parindent}}
\gdef\refis#1{\indent\hbox to 0pt{\hss#1.~}}    
\gdef\journal#1, #2, #3, 1#4#5#6{               
    {\sl #1~}{\bf #2}, #3, (1#4#5#6)}           
\def\refstylenp{                
  \gdef\refto##1{$^{(##1)}$}                                
[]
  \gdef\refis##1{\indent\hbox to 0pt{\hss##1)~}}        
  \gdef\journal##1, ##2, ##3, ##4 {                     
     {\sl ##1~}{\bf ##2~}(##3) ##4 }}
\def\refstyleprnp{              
  \gdef\refto##1{$^{(##1)}$}                                
[]
  \gdef\refis##1{\indent\hbox to 0pt{\hss##1)~}}        
  \gdef\journal##1, ##2, ##3, 1##4##5##6{               
    {\sl ##1~}{\bf ##2~}(1##4##5##6) ##3}}

\def\pr{\journal Phys. Rev., }

\def\prl{\journal Phys. Rev. Lett., }
\def\prpts{\journal Phys. Rep., }
\def\np{\journal Nucl. Phys., }
\def\pl{\journal Phys. Lett., }

\def\endreferences{\body}
\def\endpage                    
  {\vfill\eject}
\def\endpaper                   
  {\endmode\vfill\supereject}
\def\endit
  {\endpaper\end}
\def\ref#1{Ref. #1}                     
\def\Ref#1{Ref. #1}                     

\def\m@th{\mathsurround=0pt }
\font\twelvesc=cmcsc10 scaled 1200
\def\cite#1{{#1}}
\def\(#1){(\call{#1})}
\def\call#1{{#1}}
\def\taghead#1{}
\def\leaderfill{\leaders\hbox to 1em{\hss.\hss}\hfill}
\def\twiddle{\lower.9ex\rlap{$\kern-.1em\scriptstyle\sim$}}
\def\bigtwiddle{\lower1.ex\rlap{$\sim$}}
\def\gtwid{\mathrel{\raise.3ex\hbox{$>$\kern-.75em\lower1ex\hbox{$\sim$}}}}
\def\ltwid{\mathrel{\raise.3ex\hbox{$<$\kern-.75em\lower1ex\hbox{$\sim$}}}}
\def\square{\kern1pt\vbox{\hrule height 1.2pt\hbox{\vrule width 1.2pt\hskip 3pt
   \vbox{\vskip 6pt}\hskip 3pt\vrule width 0.6pt}\hrule height 0.6pt}\kern1pt}
\catcode`@=11
\newcount\tagnumber\tagnumber=0

\immediate\newwrite\eqnfile
\newif\if@qnfile\@qnfilefalse
\def\write@qn#1{}
\def\writenew@qn#1{}
\def\w@rnwrite#1{\write@qn{#1}\message{#1}}
\def\@rrwrite#1{\write@qn{#1}\errmessage{#1}}

\def\taghead#1{\gdef\t@ghead{#1}\global\tagnumber=0}
\def\t@ghead{}

\expandafter\def\csname @qnnum-3\endcsname
  {{\t@ghead\advance\tagnumber by -3\relax\number\tagnumber}}
\expandafter\def\csname @qnnum-2\endcsname
  {{\t@ghead\advance\tagnumber by -2\relax\number\tagnumber}}
\expandafter\def\csname @qnnum-1\endcsname
  {{\t@ghead\advance\tagnumber by -1\relax\number\tagnumber}}
\expandafter\def\csname @qnnum0\endcsname
  {\t@ghead\number\tagnumber}
\expandafter\def\csname @qnnum+1\endcsname
  {{\t@ghead\advance\tagnumber by 1\relax\number\tagnumber}}
\expandafter\def\csname @qnnum+2\endcsname
  {{\t@ghead\advance\tagnumber by 2\relax\number\tagnumber}}
\expandafter\def\csname @qnnum+3\endcsname
  {{\t@ghead\advance\tagnumber by 3\relax\number\tagnumber}}

\def\equationfile{%
  \@qnfiletrue\immediate\openout\eqnfile=\jobname.eqn%
  \def\write@qn##1{\if@qnfile\immediate\write\eqnfile{##1}\fi}
  \def\writenew@qn##1{\if@qnfile\immediate\write\eqnfile
    {\noexpand\tag{##1} = (\t@ghead\number\tagnumber)}\fi}
}

\def\callall#1{\xdef#1##1{#1{\noexpand\call{##1}}}}
\def\call#1{\each@rg\callr@nge{#1}}

\def\each@rg#1#2{{\let\thecsname=#1\expandafter\first@rg#2,\end,}}
\def\first@rg#1,{\thecsname{#1}\apply@rg}
\def\apply@rg#1,{\ifx\end#1\let\next=\relax%
\else,\thecsname{#1}\let\next=\apply@rg\fi\next}

\def\callr@nge#1{\calldor@nge#1-\end-}
\def\callr@ngeat#1\end-{#1}
\def\calldor@nge#1-#2-{\ifx\end#2\@qneatspace#1 %
  \else\calll@@p{#1}{#2}\callr@ngeat\fi}
\def\calll@@p#1#2{\ifnum#1>#2{\@rrwrite{Equation range #1-#2\space is bad.}
\errhelp{If you call a series of equations by the notation M-N, then M and
N must be integers, and N must be greater than or equal to M.}}\else%
 {\count0=#1\count1=#2\advance\count1
by1\relax\expandafter\@qncall\the\count0,%
  \loop\advance\count0 by1\relax%
    \ifnum\count0<\count1,\expandafter\@qncall\the\count0,%
  \repeat}\fi}

\def\@qneatspace#1#2 {\@qncall#1#2,}
\def\@qncall#1,{\ifunc@lled{#1}{\def\next{#1}\ifx\next\empty\else
  \w@rnwrite{Equation number \noexpand\(>>#1<<) has not been defined yet.}
  >>#1<<\fi}\else\csname @qnnum#1\endcsname\fi}

\let\eqnono=\eqno
\def\eqno(#1){\tag#1}
\def\tag#1$${\eqnono(\displayt@g#1 )$$}

\def\aligntag#1\endaligntag
  $${\gdef\tag##1\\{&(##1 )\cr}\eqalignno{#1\\}$$
  \gdef\tag##1$${\eqnono(\displayt@g##1 )$$}}

\def\eqalignno#1{\displ@y \tabskip\centering
  \halign to\displaywidth{\hfil$\displaystyle{##}$\tabskip\z@skip
    &$\displaystyle{{}##}$\hfil\tabskip\centering
    &\llap{$\displayt@gpar##$}\tabskip\z@skip\crcr
    #1\crcr}}

\def\displayt@gpar(#1){(\displayt@g#1 )}

\def\displayt@g#1 {\rm\ifunc@lled{#1}\global\advance\tagnumber by1
        {\def\next{#1}\ifx\next\empty\else\expandafter
        \xdef\csname @qnnum#1\endcsname{\t@ghead\number\tagnumber}\fi}%
  \writenew@qn{#1}\t@ghead\number\tagnumber\else
        {\edef\next{\t@ghead\number\tagnumber}%
        \expandafter\ifx\csname @qnnum#1\endcsname\next\else
        \w@rnwrite{Equation \noexpand\tag{#1} is a duplicate number.}\fi}%
  \csname @qnnum#1\endcsname\fi}

\def\ifunc@lled#1{\expandafter\ifx\csname @qnnum#1\endcsname\relax}

\let\@qnend=\end\gdef\end{\if@qnfile
\immediate\write16{Equation numbers written on []\jobname.EQN.}\fi\@qnend}

\catcode`@=12
\refstyleprnp
\catcode`@=11
\newcount\r@fcount \r@fcount=0
\def\refreset{\global\r@fcount=0}
\newcount\r@fcurr
\immediate\newwrite\reffile
\newif\ifr@ffile\r@ffilefalse
\def\w@rnwrite#1{\ifr@ffile\immediate\write\reffile{#1}\fi\message{#1}}

\def\writer@f#1>>{}
\def\referencefile{
  \r@ffiletrue\immediate\openout\reffile=\jobname.ref%
  \def\writer@f##1>>{\ifr@ffile\immediate\write\reffile%
    {\noexpand\refis{##1} = \csname r@fnum##1\endcsname = %
     \expandafter\expandafter\expandafter\strip@t\expandafter%
     \meaning\csname r@ftext\csname r@fnum##1\endcsname\endcsname}\fi}%
  \def\strip@t##1>>{}}

\def\citeall#1{\xdef#1##1{#1{\noexpand\cite{##1}}}}
\def\cite#1{\each@rg\citer@nge{#1}}	

\def\each@rg#1#2{{\let\thecsname=#1\expandafter\first@rg#2,\end,}}
\def\first@rg#1,{\thecsname{#1}\apply@rg}	
\def\apply@rg#1,{\ifx\end#1\let\next=\relax
\else,\thecsname{#1}\let\next=\apply@rg\fi\next}

\def\citer@nge#1{\citedor@nge#1-\end-}	
\def\citer@ngeat#1\end-{#1}
\def\citedor@nge#1-#2-{\ifx\end#2\r@featspace#1 
  \else\citel@@p{#1}{#2}\citer@ngeat\fi}	
\def\citel@@p#1#2{\ifnum#1>#2{\errmessage{Reference range #1-#2\space is bad.}%
    \errhelp{If you cite a series of references by the notation M-N, then M and
    N must be integers, and N must be greater than or equal to M.}}\else%
 {\count0=#1\count1=#2\advance\count1
by1\relax\expandafter\r@fcite\the\count0,%
  \loop\advance\count0 by1\relax
    \ifnum\count0<\count1,\expandafter\r@fcite\the\count0,%
  \repeat}\fi}

\def\r@featspace#1#2 {\r@fcite#1#2,}	
\def\r@fcite#1,{\ifuncit@d{#1}
    \newr@f{#1}%
    \expandafter\gdef\csname r@ftext\number\r@fcount\endcsname%
                     {\message{Reference #1 to be supplied.}%
                      \writer@f#1>>#1 to be supplied.\par}%
 \fi%
 \csname r@fnum#1\endcsname}
\def\ifuncit@d#1{\expandafter\ifx\csname r@fnum#1\endcsname\relax}%
\def\newr@f#1{\global\advance\r@fcount by1%
    \expandafter\xdef\csname r@fnum#1\endcsname{\number\r@fcount}}

\let\r@fis=\refis			
\def\refis#1#2#3\par{\ifuncit@d{#1}
   \newr@f{#1}%
   \w@rnwrite{Reference #1=\number\r@fcount\space is not cited up to now.}\fi%
  \expandafter\gdef\csname r@ftext\csname r@fnum#1\endcsname\endcsname%
  {\writer@f#1>>#2#3\par}}

\def\ignoreuncited{
   \def\refis##1##2##3\par{\ifuncit@d{##1}%
     \else\expandafter\gdef\csname r@ftext\csname
r@fnum##1\endcsname\endcsname%
     {\writer@f##1>>##2##3\par}\fi}}

\def\r@ferr{\endreferences\errmessage{I was expecting to see
\noexpand\endreferences before now;  I have inserted it here.}}
\let\r@ferences=\references
\def\references{\r@ferences\def\endmode{\r@ferr\par\endgroup}}

\let\endr@ferences=\endreferences
\def\endreferences{\r@fcurr=0
  {\loop\ifnum\r@fcurr<\r@fcount
    \advance\r@fcurr by 1\relax\expandafter\r@fis\expandafter{\number\r@fcurr}%
    \csname r@ftext\number\r@fcurr\endcsname%
  \repeat}\gdef\r@ferr{}\global\r@fcount=0\endr@ferences}

\let\r@fend=\endpaper\gdef\endpaper{\ifr@ffile
\immediate\write16{Cross References written on []\jobname.REF.}\fi\r@fend}

\catcode`@=12

\citeall\refto		
\citeall\ref		%
\citeall\Ref		%

\referencefile

\def\frac#1/#2{#1 / #2}

\font\titlefont=cmr10 scaled\magstep3
\def\bigtitle                      
  {\null\vskip 3pt plus 0.2fill
   \beginlinemode \doublespace \raggedcenter \titlefont}

\oneandahalfspace
\body

\bigtitle{SCALE RATIOS IN THE STANDARD
MODEL{\footnote{$^*$}{Invited Lecture at the 1994 Moriond Conference,
M\'eribel,
March 1994\hfill}}}
\author P.Ramond{\footnote{$^{**}$}{Research supported in part by the US
Department of Energy under grant DE-FG05-86-ER-40272}}
\affil Institute for Fundamental Theory, Department of Physics
 University of Florida, Gainesville, FL 32611
\vskip .3cm

\abstract
We review  the present knowledge of the Standard Model that is relevant
in formulating its possible short distance extensions. We present different
scenarios in terms of the Higgs mass, the only unknown parameter of the
model. We concentrate on the many small numbers in the model and suggest
generic methods to reproduce these numbers in terms of scale ratios,
applying see-saw like ideas to the breaking of chiral symmetries.
\endtitlepage

The Standard Model is described in terms of a mere twenty parameters,
counting Newton's constant. The challenge to theorists is to devise {\it
the} extension of the Standard Model which explain not only the number
of parameters but their values as well. Any extension will predict many
new phenomena at shorter distances. There are many candidates for
extending the Standard Model, but none have so far distinguished
themselves by reproducing the {\it values} of the parameters, not even
their multiplicity. Thus it is timely to review the types of extensions
which might generically explain the observed patterns, before plunging
in detailed models.

The Standard Model is described by three dimensionless gauge couplings
$\alpha_1$ for the hypercharge $U(1)$, $\alpha_2$ for the   weak isospin
$SU(2)$, and $\alpha_3$ for  QCD. QCD itself predicts strong CP
violation, parametrized by a fourth dimensionless parameter
$\overline\theta$.

The Higgs sector yields two parameters, a dimensionless Higgs
self-coupling, and the Higgs mass. The self coupling is expressed in
terms of the scale of electroweak breaking, which is directly
$``$measured" as the Fermi coupling. The value of the Higgs mass is the
only parameter that has not yet been determined from experiment.

The Yukawa sector of the model yields the nine masses of the elementary
fermions, which are in turn expressed as nine dimensionless Yukawa
couplings multiplied by the electroweak order parameter. This sector
also contains three mixing angles which account for interfamily decays,
and one phase which describes CP violation in these decays.

Let us start by discussing the dimensionful parameters. The most
important is Newton's constant which sets the scale. All fundamental
questions concerning dimensionful parameters should be posed in terms of
the Planck scale ($10^{-33}$ cm, or $10^{19}$  GeV). Together with the
other two fundamental constants, it sets a truly natural system of
units. The second most striking one is the value of the electroweak
order parameter, the inverse square root of the Fermi constant, in terms
of the Planck mass
$${G_F^{-1/2}\over M_{Pl}^{}}\sim 10^{-17}\ .$$
There is no satisfactory explanation for this small parameter. All
proposed extensions have strived to explain the value of this number.
One class of theories, generically
called technicolor, has proposed the existence of
strong new forces just beyond electroweak scales; this yields a natural
explanation of this parameter, but fails to explain the values of the
fermion masses. Another class of theories postulates the existence of
another type of symmetry, supersymmetry\refto{reviews}.
There, the electroweak order
parameter is related to another small parameter, the order parameter of
supersymmetry breaking. This may not seem very economical, but it is
remarkable that supersymmetry breaking automatically generates
electroweak breaking\refto{trigger}
 in a wide class of theories. Thus it appears that
there is nothing gained nor lost. The ideas of technicolor can then be
successfully applied to supersymmetry breaking, by means of gaugino
condensation, without the problem of fermion masses. Thus many believe
that supersymmetry provides the best hope for explaining both the
electroweak breaking scale and the value of the fermion masses.

All quark and charged lepton masses break weak isospin by half a unit,
along $\Delta I_W={1\over 2}$, with the same quantum numbers as the
electroweak order parameter, which gives the W-boson its mass. It is
thus natural to form the dimensionless ratio
$${m_t\over M_W}\sim {\cal O}(1)\ ,$$
which has a natural value. However there are other quark masses for
which these ratios are much smaller,
$${m_{u,d}\over M_W}\sim 10^{-4}
\ ;\qquad
{m_s\over M_W}\sim 10^{-3}\ ;\qquad {m_c\over M_W}\sim 10^{-2}\ ;
\qquad {m_b\over M_W}\sim .05\ .$$
Similarly for the charged leptons
$${m_e\over M_W}\sim {\cal O}(10^{-5})\ ;\qquad
{m_\mu\over M_W}\sim {\cal O}(10^{-3})\ ;\qquad {m_\tau
\over M_W}\sim .02\ ,$$
range from the tiny to the small.

The neutrino masses are predicted to be exactly zero in the standard
model only because of the global lepton number symmetries. However
neutrino masses, if they were to be non-zero, would break weak isospin
by one unit, that is have $\Delta I_W=1$ values. Experimental limits on
neutrino masses indicate that they are at most extremely small. For
instance,
$${m_{\nu_e}\over M_W}< 10^{-17}\ .$$
Interestingly, this is reflected by the fact that weak isospin shows no
sign of having been broken in that direction. We should mention that the
masslessness of the photon and the gluons is deemed natural since
protected by a gauged symmetry.

The values of the three gauge parameters are known to great accuracy.
Because of endemic problems associated with strong QCD, that coupling is
the least well known.

Given all these parameters, we can extrapolate the Standard Model to
shorter distances, using the renormalization group. The most interesting
effect occurs in the extrapolation of the three gauge couplings. We
normalize the hypercharge coupling as if it were part of a non-Abelian
group in which the standard model groups fit snuggly ($SU(5),\ SO(10),\
E_6^{}$\refto{gut}).
We find that the hypercharge and weak isospin couplings meet
at a scale of $10^{13}$ GeV, with a value $\alpha^{-1}\approx 43$. We
also find that at that scale, the QCD coupling is much larger,
$\alpha_3^{-1}\approx 38$. Thus, although the quantum numbers indicate a
possible unification into a larger non-Abelian group, the gauge coupling
do not follow suit in this naive extrapolation. Historically of course,
before the couplings were known to this accuracy, it was believed that
all three did indeed unify in the ultraviolet. In any case, the lack of
observed proton decay restricts the scale of unification to above
$10^{16}$ GeV. One can still say that in the ultraviolet, the values of
these couplings is less disparate than at experimental scales.
Similarly, nothing spectacular occurs to the Yukawa couplings. For
instance, the botton quark and $\tau$ lepton Yukawa couplings meet
around $10^9$ GeV, but diverge in the deeper ultraviolet.

The situation is potentially more interesting in the Higgs sector
because of the renormalization group behavior of the Higgs self
coupling\refto{sher}. We can consider two cases, depending on the value of the
Higgs
mass.

If the Higgs mass is below $150$ GeV, the self-coupling turns negative
at shorter diarances. This results in an unbounded potential, and
instability of the standard model beyond the scale at which it changes
sign. For example, using the recently measured value of the top quark
mass, we find that a Higgs mass of $120$ GeV would mean instability
setting in at 1 TeV. In such case, new particles with masses
commensurate with that scale must exist to stabilize the theory. This is
exactly what happens in the supersymmetric extension of the Standard
Model. One may envisage other stabilizing schemes without supersymmetry,
but it is just the most tractable.

If the Higgs mass is above $200$ GeV, the self-coupling rises
dramatically towards its Landau pole at a relatively low energy scale.
This only means that we lose perturbative control over the theory. It
sets an upper bound on the Higgs mass since there is no evidence of
strong coupling at our scale. This is called the triviality limit
because, looked at from the other side, it drives the self-coupling to
zero in the infrared. However we know that the coupling is {\it not}
zero for the standard model; thus strong coupling must happen. In all
likelihood, this means that the Higgs is a composite; an example of this
view is the technicolor scenario where the Higgs is a condensate of
techniquarks.

Within  a tiny range of intermediate values for the Higgs mass, the
instability and triviality bounds are pushed to scales beyond the Planck
length. In this case, there is no Standard Model prediction of new
physics, except for the usual caveats associated with quantum gravity.
Then we should view the Planck mass as the physical cut-off of any
theory at lower energies. It is instructive to see what happens to the
various Standard Model parameters in terms of the Planck cut-off.

The most striking behavior is that the renormalization of the Higgs mass
is proportional to the cut-off itself. This does not make it natural to
envisage a light Higgs with such an enormous renormalization.
Thus even if the Higgs mass does not demand new physics below Planck
mass, it makes for a pretty {\it ad hoc} theory. We can contrast the
situation with fermion masses. Their dependence on the cut-off is only
logarithmic. The reason is that a fermion mass is natural in the sense
that by setting it to zero, one gains a chiral symmetry that is
respected by quantum corrections. This allows for a protection mechanism
which results in a weak cut-off dependence.

Supersymmetry avoids the naturalness problem in the following way: it
links any fermion to a boson of the same mass, so that in the limit of
exact supersymmetry, the boson mass is also protected by the chiral
symmetry hitherto associated with the fermion. This is enough protection
to assure, even after supersymmetry breaking, a mass for the Higgs that
is commensurate with the scale of supersymmetry breaking.

This might seem to be small progress, since a new symmetry has been
introduced to relax the strong cut-off dependence. That new symmetry has
to be broken itself at a small scale. Indeed, in order to reproduce the
value of Fermi's constant, we must be able to obtain
$${V^{}_{SUSY}\over M_{Pl}}\sim 10^{-15}\ ,$$
where $V^{}_{SUSY}$ is the supersymmetry breaking order parameter.
Assume for a moment we know how to do this, and see if we have gained
anything.

The first thing is that the gauge couplings seem to be much closer to
unification, and at a scale not  invalidated by proton decay bounds. One
finds that the hypercharge and weak isospin couplings meet at a scale of
the order of $10^{16}$ GeV, with a value $\alpha^{-1}\approx 25$. In
this case, however, the QCD coupling is much closer to, if not right on the
same value\refto{unification}.
It may still be a shade higher than the others, with
$(\alpha^{-1}-\alpha_3^{-1})\le 1.5$.

The second thing is that with this value, and suitable boundary
conditions at or near Planck mass, the renormalization group drives one
of the Higgs masses to imaginary values in the infrared. This in turns
triggers electroweak breaking, made possible only because of the
large top quark mass.

It is significant that the extension to supersymmetry yields a model
with no couplings that blow up below Planck mass. For example, the Higgs
self-coupling is replaced by gauge couplings which are ultraviolet-tame.
However, the Higgs mass is not arbitrarily high in the minimal
extension. At tree-level, it is predicted to be below the Z-mass, but it
suffers large radiative corrections due to the top Yukawa coupling,
raising it above the Z, but not by an arbitrarily large amount\refto{gordy}.

This general scheme allows us to study the pattern of fermion masses at
these shorter distances. It is interesting that there are more
regularities with supersymmetry than without. For instance, the bottom
quark and $\tau$ masses seem to unify at or around $10^{16-17}$
GeV\refto{btau}.

As we have seen, most of the parameters yet to be explained are to be
found in the Yukawa sector. With supersymmetry, the observed pattern of
Yukawa couplings can be extrapolated all the way to or near Planck
length. The hope is that at that scale, where things are supposed to be
simpler, there might emerge some patterns not recognized at lower
energies.

The most striking aspect of the fermion masses is that only the third
family has sizeable masses. Thus it is natural to consider theories
where the Yukawa matrices are simply of the form
$${\bf Y}_{u,d,e}=\pmatrix{0&0&0\cr 0&0&0\cr
0&0&y^{}_{t,b,\tau}\cr}\ .$$
These matrices imply an enormous global chiral symmetry in each sector
of the $U(2)_L\times U(2)_R$. There is of course the hierarchy between
the bottom and top quark masses which must also be explained. In the
$N=1$ model, it is related to another parameter which comes from the
Higgs sector, the ratio of the {\it vev} of the two Higgs. We do not
concern ourselves with it here. Thus the question of interest is really
why are the other two families so light? In order to gain some
perspective on this question, let us examine one well-known case in
which small numbers are naturally generated, the see-saw
mechanism\refto{gmrsy}.

The Standard Model neutrino Majorana mass matrix is zero. How do we fill
the zeros, which are protected by lepton number conservation? They can
be filled only if lepton symmetry is broken.

What happens in the see-saw mechanism is that the usual neutrinos are
mixed with new electroweak singlet neutrinos. This gives them the same
lepton numbers. Then the lepton numbers are broken by giving these
neutrinos a mass $M$, which breaks lepton number at the same scale $M$.
Upon diagonalization, this generates an entry in the mass matrix which
is depressed from its expected value by the ratio of scale ${m\over M}$,
where $n$ is the typical electroweak scale.

Let us analyze the charged Yukawa matrices in the same way. The zeros of
the Yukawa matrices are protected by chiral symmetries. Thus we first
couple the massless fermions with fermions with similar quantum numbers.
This shares the chiral symmetries with the new fermions. Then we assume
these new fermions, being charged have a vector-like partner (this
differs from the neutral sector), and that they can acquire $\Delta
I_W=0$ mass at a new scale M. This mass breaks the chiral symmetry. Upon
diagonalization, this fills some of the entries.

Consider a generic model with $3$ left-handed fields $f^{}_i,f^{}_3$.
Assume that the only tree level Yukawa involving the chiral fields is of
the form $f^{}_3f^{}_3h$, where $h$ is a Higgs which can break
electroweak symmetry. In the absence of any other couplings, this leaves
us with a left-handed $U(2)$ symmetry acting on $f^{}_{1,2}$. Thus if
this symmetry is not broken, these fields will forever remain massless,
at least in perturbation theory. We have to find a way to break this
chiral symmetry.

To do this, let us add to the model $N$ vector-like families
$F^{}_a\oplus\overline F^{}_b$. After marrying off the left and right
handed fields, we still have three chiral families. Such a situation
generically arises in superstring compactifications where vector-like
particles are readily available. The importance of these fields is that
they can be used to break the chiral symmetry. First we observe that
they can have $\Delta I_W=0$ masses which do not break electroweak
symmetry, of the form $M_{ab}\overline F^{}_aF^{}_b$. These terms break
the chiral symmetries associated with the vector-like families. In order
to relate the two types of chiral symmetries, we must couple these to
the $f_i$. There are two types of such terms. The first is itself
vector-like, and can occur at the large scale: $f_{i,3}^{}\overline
F^{}_a$. The second type is of the form $f^{}_{i,3}F^{}_ah$, and breaks
the electroweak symmetry, of the same type found in the see-saw
mechanism. We can of course consider chiral operators of the form
$F^{}_aF^{}_b$, and their conjugates, as well. However these might yield
extra light particles in the spectrum, since these operators have
electroweak quantum numbers. Thus we do not include them.

For instance, with one vector-like family, we may consider a Lagrangian
of the form
$$f^{}_3f^{}_3h+(f^{}_1+f^{}_2+f^{}_3+f^{}_4)\overline FH\ ,$$
where $h$ is the electroweak breaking Higgs, and $H$ is in the $\Delta
I_W=0$ sector. The fermion fields are all left-handed and refer to
families. We have seven fields, and five terms in the Lagrangean,
leaving us with two symmetries, which are both broken when $h$ and $H$
get vacuum values. Upon diagonalization of the mass matrix, we find two
tree-level zero eigenvalues, but they are unprotected by chiral
symmetries, and will be radiatively corrected. This model can be easily
implemented in $SO(10)$ and $E_6$. However the symmetries do not forbid
couplings such as $f_1^{}f^{}_3h$; one has to appeal to supersymmetry to
explain the naturalness of these zeros.

It is a matter of model building to come up with specific arrays of
vector-like particles which reproduce the family hierarchy. It is not
easy to come up with such models, in the absence of extra symmetries.

{}From the point of view of a low energy effective theory, where the
effect of the massive vector-like particles have been integrated out,
the zeros will be filled ny non-renormalizable effective operators of
the form
$$f^{}_if^{}_jh\left({K\over M}\right)^{n_{ij}}\ ,$$
where $h$ is the usual Higgs field, and $K$ is a combination of Higgs
doublets which can get non-zero vacuum value, and $M$ is a large mass.
The exponents $n_{ij}$ may be determined by symmetry. In order to
produce a small coefficient, the $i$th and $j$th fermions need to go
through a number of intermediate steps to interact. The larger the
number steps, the larger $n_{ij}$, and the smaller the entry in the
effective Yukawa matrix. This approach was advocated long ago by
Froggatt and Nielsen\refto{FN}.

One may take the point of view that these non-renormalizable operators
come from physics beyond the Planck scale, in which case, the question is
relegated to one of classifying the possible non-renormalizable
operators, without having to say how they are generated.

Clearly much work needs to be done before any successfull model of this
type is devised, but we believe that this is a correct framework to
analyze the Yukawa patterns.

I wish to thank the organizers of this Moriond Encounter for their kind
hospitality, and the invigorating intellectual atmosphere they provided.
\vskip 1cm
\references

\refis{FN} C.~Froggatt and H.~B.~Nielsen \np B147, 277, 1979.

\refis{reviews}
For reviews, see H.~P.~Nilles,  \prpts 110, 1, 1984 and
H.~E.~Haber and G.~L.~Kane, \prpts 117, 75, 1985.

\refis{trigger}
L.~E.~Ib\'a\~nez and  G.~G.~Ross,
\journal Phys.~Lett., 110B, 215, 1982;
K.~Inoue, A.~Kakuto, H.~Komatsu, and S.~Takeshita,
\journal Prog. Theor. Phys., 68, 927, 1982;
L.~Alvarez-Gaum\'e, M.~Claudson, and M.~Wise,
\np B207, 16, 1982;
J.~Ellis, J.~S.~Hagelin, D.~V.~Nanopoulos, and
K.~Tamvakis,
\journal Phys.~Lett., 125B, 275, 1983.

\refis{gut}
J.~C.~Pati and A.~Salam,
\pr D10, 275, 1974;
H.~Georgi and S.~Glashow,
\prl 32, 438, 1974;
H.~Georgi, in {\it Particles and Fields-1974}, edited by C.E.Carlson,
AIP Conference Proceedings No.~23 (American Institute of Physics,
New York, 1975) p.575;
H.~Fritzsch and P.~Minkowski,
\journal Ann.~Phys.~NY, 93, 193, 1975;
F.~G\" ursey, P.~Ramond, and P.~Sikivie,
\pl 60B, 177, 1975.

\refis{btau} H.~Arason, D.~J.~Casta\~no, B.~Keszthelyi, S.~Mikaelian,
E.~J.~Piard, P.~Ramond, and B.~D.~Wright,
\prl 67, 2933, 1991;
A.~Giveon, L.~J.~Hall, and U.~Sarid,
\pl 271B, 138, 1991.

\refis{gordy} G.~L.~Kane, C.~Kolda, and J.~D.~Wells,
\prl 70, 2686, 1993.

\refis{sher} M.~Sher, \prpts 179, 273, 1989.

\refis{unification}
U.~Amaldi, W.~de Boer, and H.~Furstenau,
\pl B260, 447, 1991;
J.~Ellis, S.~Kelley and D.~Nanopoulos,
\pl 260B, 131, 1991;
P.~Langacker and M.~Luo,
\pr D44, 817, 1991.

\refis{gmrsy}M. Gell-Mann, P. Ramond, and R. Slansky in Sanibel
Talk,
CALT-68-709, Feb 1979, and in {\it Supergravity} (North Holland,
Amsterdam 1979). T. Yanagida, in {\it Proceedings of the Workshop
on Unified Theory and Baryon Number of the Universe}, KEK, Japan,
1979.

\endreferences\end